\documentclass[aps,prl,twocolumn,showpacs,superscriptaddress]{revtex4}

\usepackage{graphicx}  
\usepackage{dcolumn}   
\usepackage{bm}        
\usepackage{amssymb}   
\usepackage{amsmath}
\usepackage{units}
\usepackage{times}
\usepackage{textcomp}
\begin{document}

\title{Surface band structure of $\text{Bi}_{1-x}\text{Sb}_{x}$(111) }

\author{Hadj M. Benia}
\email[Corresponding author; electronic address:\ ]{h.benia@fkf.mpg.de}
\affiliation{Max-Planck-Institut f\"ur
Festk\"orperforschung, 70569 Stuttgart, Germany}
\author{Carola Stra\ss er}
\affiliation{Max-Planck-Institut f\"ur Festk\"orperforschung,
70569 Stuttgart, Germany}
\author{Klaus Kern}
\affiliation{Max-Planck-Institut f\"ur Festk\"orperforschung,
70569 Stuttgart, Germany} \affiliation{Institut de Physique de la Mati{\`e}re Condens{\'e}e, Ecole Polytechnique
F{\'e}d{\'e}rale de Lausanne, 1015 Lausanne, Switzerland}
\author{Christian R. Ast}
\affiliation{Max-Planck-Institut f\"ur Festk\"orperforschung,
70569 Stuttgart, Germany}

\date{\today}

\begin{abstract}
Theoretical and experimental studies agree that $\text{Bi}_{1-x}\text{Sb}_{x}$ ($0.07 \leq x \leq 0.21$) to be a three-dimensional topological insulator. However, there is still a debate on the corresponding $\text{Bi}_{1-x}\text{Sb}_{x}$(111) surface band structure. While three spin polarized bands have been claimed experimentally, theoretically, only two surface bands appear, with the third band being attributed to surface imperfections. Here, we address this controversy using angle-resolved photoemission spectroscopy (ARPES) on $\text{Bi}_{1-x}\text{Sb}_{x}$ films. To minimize surface imperfections, we have optimized the sample growth recipe. We have measured the evolution of the surface band structure of $\text{Bi}_{1-x}\text{Sb}_{x}$ with $x$ increasing gradually from $x = 0$ to $x = 0.6$. Our ARPES data show better agreement with the theoretical calculations, where the system is topologically non-trivial with two surface bands.
\end{abstract}

\maketitle

Topological insulators (TIs) are characterized by ungapped and protected edge/surface states that render the surface metallic. These states exhibit a non-trivial topology that imposes an odd number of crossings with the Fermi level \cite{fu_topological_2007,hasan_colloquium:_2010}. The first angle-resolved photoemission spectroscopy (ARPES) data showing the non-trivial topology have been measured on the (111) surface of the semiconducting phase of a $\text{Bi}_{1-x}\text{Sb}_{x}$ single crystal for $x=0.1$ \cite{hsieh_topological_2008}. Its experimental band structure is similar to pure Bi(111) especially at the $\overline{\Gamma}$-point where two spin-polarized surface bands emerge from the bulk valence band continuum \cite{ast_electronic_2003,ast_high-resolution_2004,hofmann_surfaces_2006,hsieh_topological_2008}. However, the band structure around the $\overline{\textit{M}}$-point is still controversial. While experimentally the topological character has been claimed by the presence of a third spin polarized band (absent in Bi(111)) and correspondingly five crossings with the Fermi level \cite{hsieh_topological_2008,hsieh_observation_2009,nishide_direct_2010,nakamura_topological_2011}, theoretically the number of crossings is also odd, but the configurations of the topological surface states do not include a third surface band \cite{teo_surface_2008,zhang_electronic_2009}. This additional band has been ascribed to result from imperfect surfaces \cite{zhang_electronic_2009}.

Here, we experimentally examine the above debate using ARPES on $\text{Bi}_{1-x}\text{Sb}_{x}$ films. We optimized the \textit{in situ} film growth method to minimize surface imperfections. We were able to control Sb concentration from $x = 0$ to $x = 0.6$ as Sb content is a critical parameter to determine the electronic properties of the $\text{Bi}_{1-x}\text{Sb}_{x}$ alloy. In order to have an overall view on the surface band structure, we tracked the evolution of the surface states not only near the $\overline{\text{M}}$-point but also around the $\overline{\Gamma}$-point. The ARPES results show on one hand a gradual evolution of the surface band structure from Bi(111) towards Sb(111), attesting precise control of the Sb content. On the other hand, we show that the third surface band could not be detected in the topological regime. Still, the corresponding surface band structure remains topological in accordance with the theory.

\begin{figure*}
\centerline{ \includegraphics[width = 0.95\textwidth]{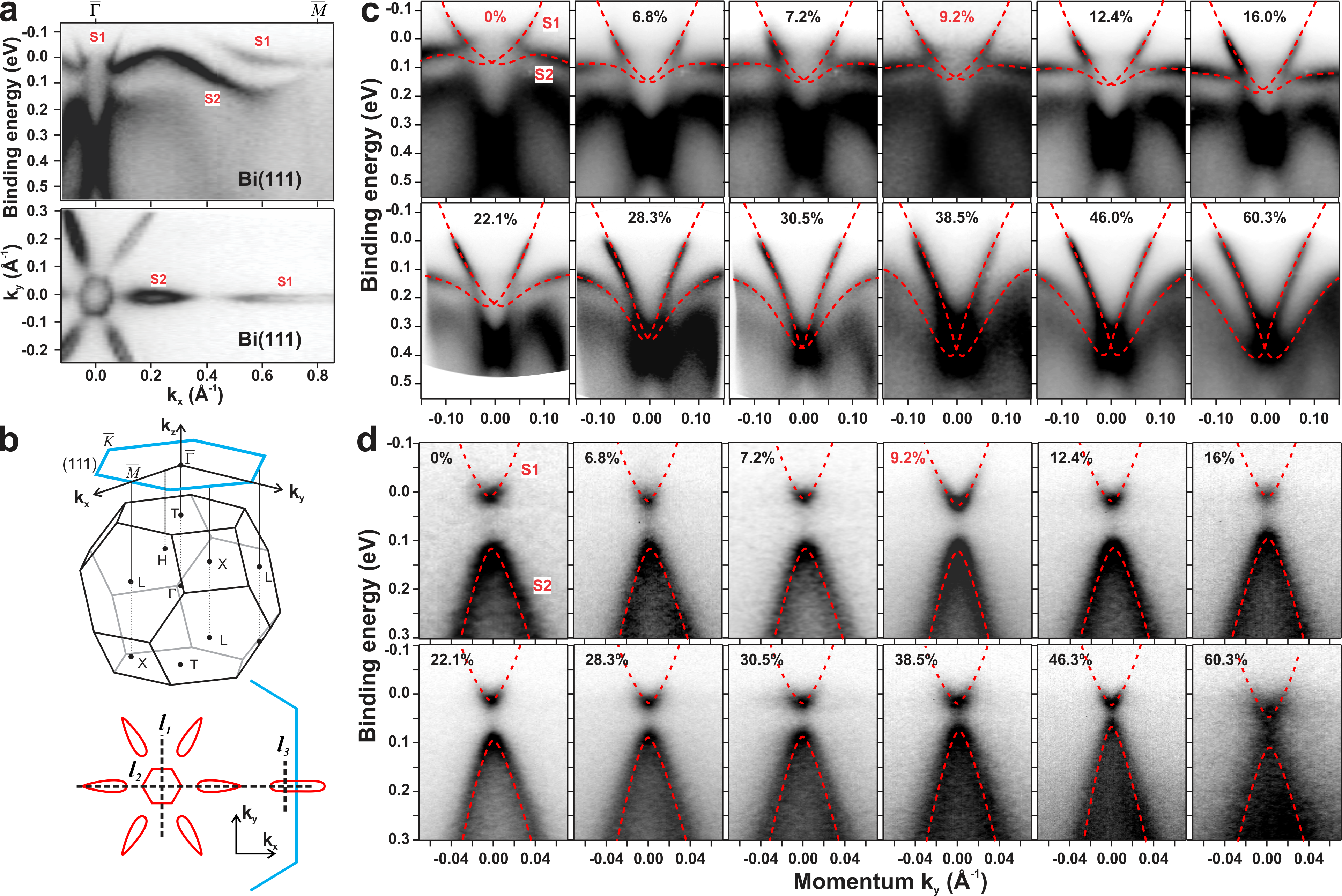}}
\caption{(a) Surface state band structure of Bi(111) along the $\overline{\Gamma\textit{M}}$ direction (line $l_2$, Fig.\ \ref{fig:ARPESevo}-b) and the corresponding Fermi surface recorded at 300\,K. (b) Top: schematic representation of the 3D and 2D Brillouin zones of Bi crystal and its (111) surface. Bottom: schematic representation of the Fermi surface of Bi(111). The dashed lines denoted by $l_1$, $l_2$, and ${l_3}$ represent the different momentum directions along which the band structures have been recorded. $l_3$ corresponds to $k_x=0.667$ $\text\AA^{-1}$, i.e, at a polar angle of $18${\textdegree} using a He I light source. (c) and (d) Evolution of the experimental surface band structure of $\text{Bi}_{1-x}\text{Sb}_{x}$(111) as a function of $x$ near $\overline{\Gamma}$ along $\overline{\Gamma\textit{K}}$ (line $l_1$, Fig.\ \ref{fig:ARPESevo}-b) and near $\overline{\textit{M}}$ along line $l_3$ in \ref{fig:ARPESevo}-c, respectively. The red dashed lines on the surface states are guides to the eye. The Sb concentration $x$ is indicated for measurements performed at 100 K in black and at 300 K in red.} \label{fig:ARPESevo}
\end{figure*}

For the preparation of $\text{Bi}_{1-x}\text{Sb}_{x}$ samples, we have adopted an optimized growth procedure that produces high quality surfaces yielding sharp experimental band structure for Sb concentrations $0 \leq x \leq 0.6$. First, a 30\,nm thick pure Bi(111) film is grown on a Si(111)-7$\times$7 substrate. The sample is post-annealed at 500\,K. As shown in Fig.\ \ref{fig:ARPESevo}-a, this growth method gives rise to a sharp and intense ARPES structure even at room temperature. On this buffer layer, we grow 120\,nm thick $\text{Bi}_{1-x}\text{Sb}_{x}$(111) films. Bi and Sb are simultaneously deposited from Knudsen (effusion) cells.  The Sb concentration has been determined from x-ray photoemission (XPS) spectra. An atomic sensitivity factor ratio $(K= 1.23)$ has been used to find the Sb to Bi concentration ratio. The ratio $K$ was determined independently by energy dispersive x-ray spectroscopy (EDX). The ARPES measurements were performed with a hemispherical SPECS HSA3500 electron analyzer characterized by an energy resolution of about 10\,meV. Monochromatized He I (21.2\,eV) radiation was used as a photon source. The sample was measured either at 100\,K or at room temperature to follow the dispersion of the surface states above the Fermi level.

\begin{figure}
\centerline{ \includegraphics[width = 1.0\columnwidth]{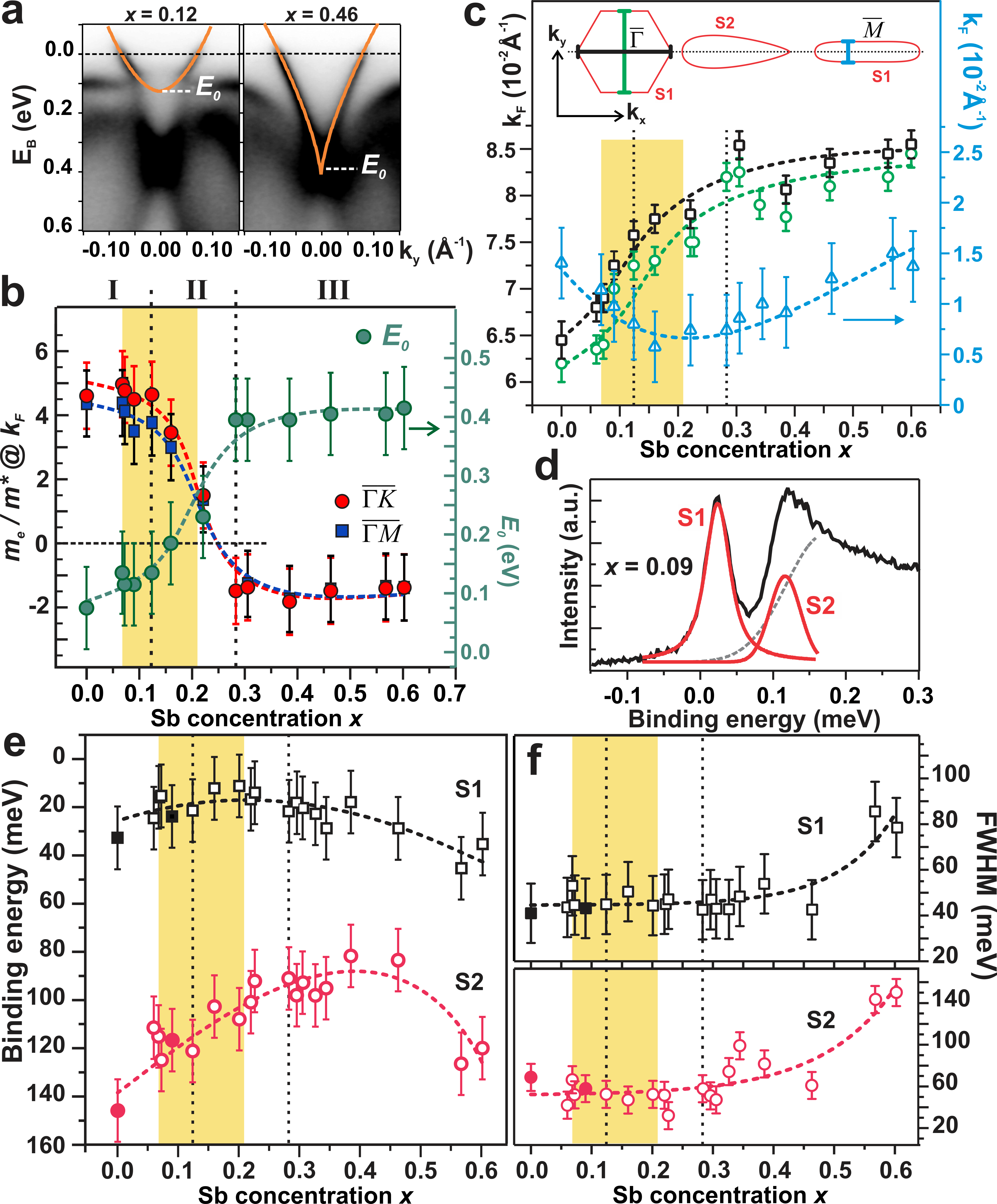}}
\caption{Evolution of different parameters as a function of Sb concentration. (a) Surface state band structure along the $l_1$ line (Fig.\ \ref{fig:ARPESevo}-b) of $\text{Bi}_{0.88}\text{Sb}_{0.12}$ and $\text{Bi}_{0.44}\text{Sb}_{0.46}$ with the corresponding fitting curves (orange) to the dispersion of $S1$. (b) Inverse effective mass $1/m^\ast$ at $k_F$ obtained from the fit of $S1$ dispersion around $\overline{\Gamma}$. Estimated binding energy $E_0$ of the $S1$ and $S2$ crossing at the $\overline{\Gamma}$-point. (c) Fermi vector $k_F$ along $k_x$ (squares) and $k_y$ (circles) of the electron pocket around the $\overline{\Gamma}$-point and of the second electron pocket near $\overline{\textit{M}}$ along the $l_3$ line (blue triangles). The sketch on the top of the graph representing a Fermi surface of Bi(111) along $\overline{\Gamma{\textit{M}}}$ is a graphical legend of the graph. (d) Example of the fit of EDCs at $k_y$ = 0 \AA$^{-1}$ along the $l_3$ line from $\text{Bi}_{0.91}\text{Sb}_{0.09}$ measured at 300 K. The fit curves are $S1$ and $S2$ (red) and background contributions (dashed-gray). (e) Energy peak positions of $S1$ and $S2$ in the EDCs at $k_y$ = 0 \AA$^{-1}$ of the band structure recorded along the $l_3$ line, (f) corresponding extracted line widths of $S1$ and $S2$ peaks. The yellow zone indicates the TI region. The solid symbols in (e) and (f) indicate room temperature data points.}\label{fig:fits}
\end{figure}

Figure \ref{fig:ARPESevo}-c presents the experimental surface band-structure of $\text{Bi}_{1-x}\text{Sb}_{x}$ along the $l_1$ line (Fig.\ \ref{fig:ARPESevo}-b) as Sb concentration increases progressively from $x = 0$ to $x = 0.6$. In all panels, the surface states $S1$ and $S2$ appear sharp and intense. The broad features below the surface states around the $\overline{\Gamma}$-point are surface resonances appearing within the projected bulk valence band \cite{ast_electronic_2003,ast_high-resolution_2004,hofmann_surfaces_2006,sugawara_fermi_2006}. Figure \ref{fig:ARPESevo}-c nicely shows a gradual evolution of the surface band structure near the $\overline{\Gamma}$-point going from a pure Bi(111)-like to a pure Sb(111)-like band structure. As is known from the Bi(111) surface band structure, $S1$ and $S2$ lose spectral weight near the $\overline{\Gamma}$-point since they disperse into the projected bulk valence band \cite{ast_electronic_2003,ast_high-resolution_2004,hofmann_surfaces_2006}. The crossing of the spin-split bands $S1$ and $S2$ cannot, therefore, be discerned. The red dashed lines crossing at the $\overline{\Gamma}$-point in Fig.\ \ref{fig:ARPESevo}-c are an extrapolation of the experimental dispersion of $S1$ and $S2$ based on a theoretically calculated band structure of Bi(111) \cite{koroteev_strong_2004}.

\begin{figure}
\centerline{\includegraphics[width = 0.99\columnwidth]{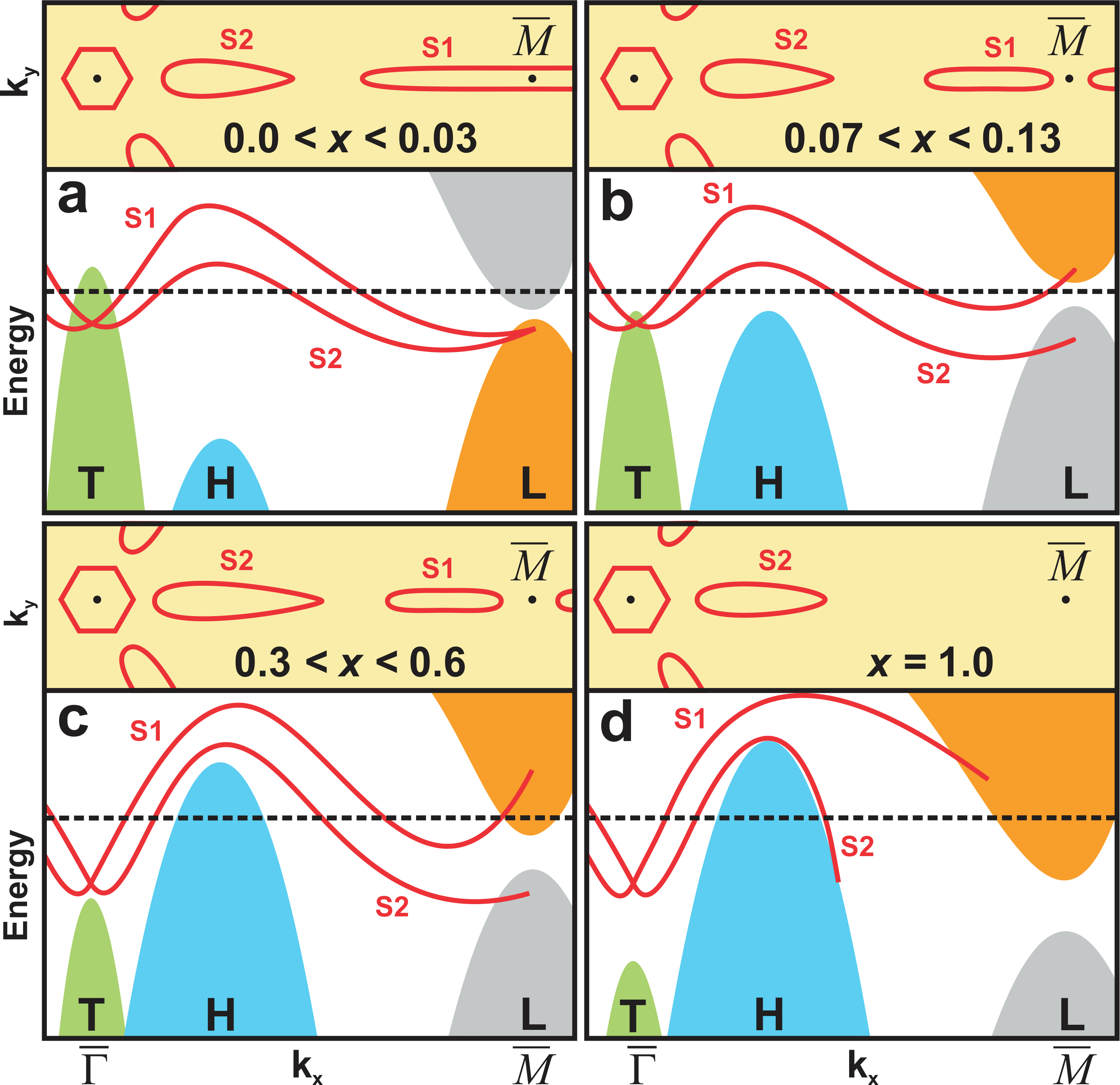}}
\caption{Schematic representation of the main phases during the evolution of the band structure of the $\text{Bi}_{1-x}\text{Sb}_{x}$(111) surface with increasing Sb concentration. The top panels are the corresponding Fermi surface representations.} \label{fig:Scenario}
\end{figure}

The surface band structure of Bi(111) in Fig.\ \ref{fig:ARPESevo}-c remains almost unchanged when increasing $x$ from $0$ to $0.16$. For $x>0.22$, a deformation of the $S1$ dispersion occurs: The effective mass $m^\ast$ of $S1$ changes sign at $x=0.28$. In order to analyze the evolution of $m^\ast$ of $S1$ as a function of Sb concentration, we fitted its dispersion with a symmetric power function. Figure \ref{fig:fits}-a presents different fits to $S1$ at two different concentrations. The fit details are summarized in the supplemental material \cite{Epaps}. In Fig.\ \ref{fig:fits}-b, we plot the extracted evolution of $1/m^\ast$ at the Fermi level. Three main Sb-concentration regions can be distinguished. They are denoted as regions I to III and delimited by dotted vertical lines in Fig.\ \ref{fig:fits}. In region I ($0 < x\lesssim 0.13$), the $m^\ast$ anisotropy along $\overline{\Gamma\textit{K}}$ and $\overline{\Gamma\textit{M}}$ is due to the hexagonal warping of the electron pocket formed by $S1$ \cite{ast_fermi_2001,hofmann_surfaces_2006}. In region II ($0.13 \lesssim x\lesssim 0.28$), the sign of $m^\ast$ changes, so that at $x \approx 0.25$, the band dispersion is linear. In this region, the transition of the dispersion of the surface states from a Bi(111)-like configuration (region I) to an Sb(111)-like configuration (region III)  takes place. We attribute this change to the shifting bulk bands near the T and H points \cite{lenoir_transport_1996,tang_phase_2012-1}. While the band at the T point shifts to higher binding energy ($E_B$) for increasing $x$, the band at the H point shifts towards the Fermi level (see Fig.\ \ref{fig:Scenario}).

The evolution of the experimental band structure of the surface states along the $l_3$ line near the $\overline{\textit{M}}$-point is shown in Fig.\ \ref{fig:ARPESevo}-d as a function of Sb content. The bands are sharp and intense, but seem to not deviate much from the pure Bi(111) band dispersion in the entire concentration range, except at $x=0.6$, where they become faint and washed-out. This broadening and intensity loss is not related to the surface quality of the sample, since the bands are very sharp around $\overline{\Gamma}$ (Fig.\ \ref{fig:ARPESevo}-c). We have analyzed the energy distribution curves (EDC) at $k_y=0$ of the band structures in Fig.\ \ref{fig:ARPESevo}-d by fitting a Voigt function to the spectral features of the $S1$ and $S2$ bands (Fig.\ \ref{fig:fits}-d).
The energy positions of $S1$ and $S2$ (Fig.\ \ref{fig:fits}-e) evolve smoothly as $x$ increases. The binding energy of $S1$ varies from $E_B\approx 30$ meV for pure Bi, goes to a minimum of $E_B\approx 15$ meV for $x\approx0.2$, then increases again and reaches $E_B\approx 40$ meV at $x\approx0.6$. Similarly, the $S2$ energy position decreases slowly from $E_B\approx 140$ meV at $x=0$ to a minimum $E_B\approx 90$ meV for an Sb content of $x\approx0.4$ and then increases again to $E_B\approx 120$ meV at $x\approx0.6$. We note that the TI region does not constitute any special stage in the evolution of $S1$ and $S2$ energy positions \cite{hsieh_topological_2008,lenoir_transport_1996,tang_phase_2012-1,nishide_direct_2010}. In order to check for any hidden phenomena within the linewidth of the two surface bands, we analyzed the linewidth (FWHM) for each spectral feature. The extracted evolution of the linewidth for $S1$ and $S2$ is also smooth as shown in Fig.\ \ref{fig:fits}-f. The band $S1$ is characterized by a smaller linewidth than $S2$, which is most likely due to a longer quasiparticle lifetime near the Fermi level. For $x=0$, $S1$ has a linewidth of about $40$ meV, which stays almost constant with increasing Sb content until $x=0.4$. For $x>0.4$, it increases comparatively rapidly to around $80$ meV at $x=0.6$. The linewidth of $S2$ follows the behavior of $S1$ with constant values around $55$\,meV until $x\approx0.4$, after which it increases to about $140$ meV at $x=0.6$.

\begin{figure*}
\centerline{ \includegraphics[width = 0.99\textwidth]{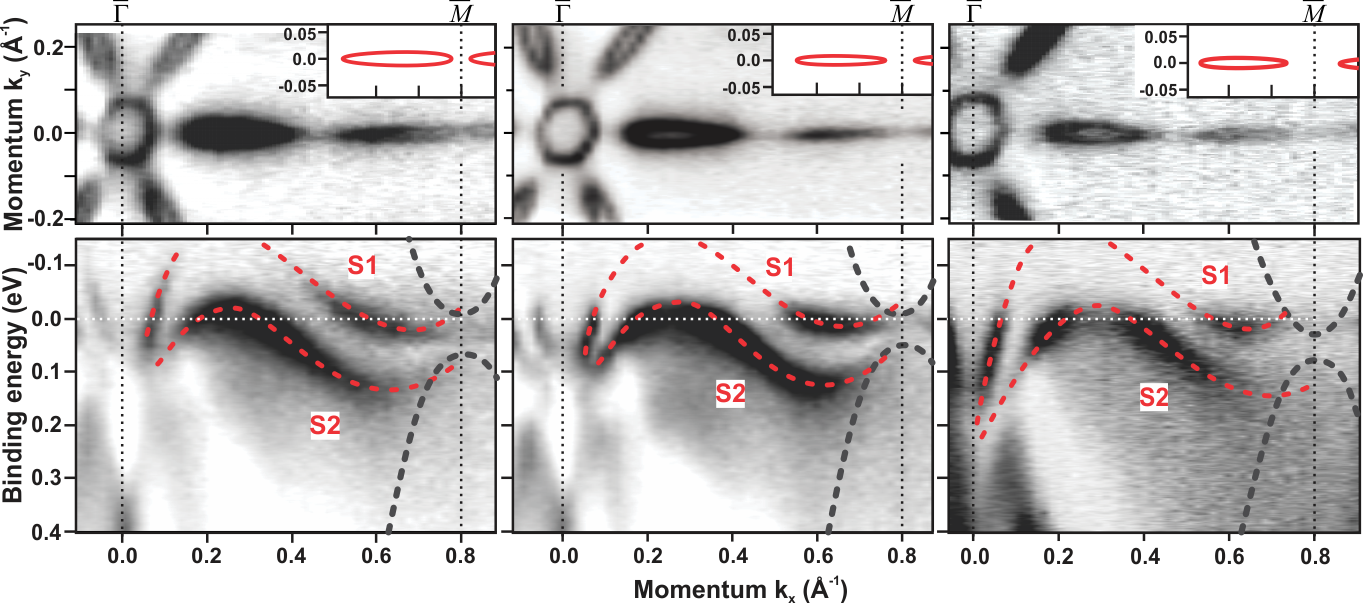}}
\caption{(Top) Fermi surface and (bottom) corresponding experimental surface band structure along the $\overline{\Gamma{\textit{M}}}$ direction  of $\text{Bi}_{0.91}\text{Sb}_{0.09}$ (300 K), $\text{Bi}_{0.89}\text{Sb}_{0.11}$ (100 K), and $\text{Bi}_{0.7}\text{Sb}_{0.3}$ (100 K). The insets indicate the closing electron-pocket contour before the $\overline{\textit{M}}$-point.} \label{fig:GamaMbandstrcuture}
\end{figure*}

We note that the linewidth of $S1$ does not show any additional broadening within the TI region. This observation is different from previous ARPES results that indicate an additional $S1$ broadening \cite{hirahara_topological_2010,nakamura_topological_2011,nishide_direct_2010,guo_evolution_2011}. Those ARPES measurements show the presence of a weak contribution to the linewidth of $S1$ for $x\approx 0.13, 0.16, 0.17$, and $0.21$ \cite{nishide_direct_2010,nakamura_topological_2011,hirahara_topological_2010,guo_evolution_2011}. An extra broadening of $S1$ could agree with the appearance of a third spin-polarized surface band ($S3$), which would result in five crossings with the Fermi level proving the non-trivial topology classification of the system as indicated in Ref. \cite{hsieh_topological_2008,hsieh_observation_2009}. However, the expected energy position of $S3$ is about $45$ meV from $S1$ at $k_x = 0.67$ $\text\AA^{-1}$ \cite{hsieh_topological_2008,hsieh_observation_2009}. With a linewidth of $S1$ of about $40$ meV, a shoulder near the $S1$ peak in the EDC curves corresponding to $S3$ should be observable. Here, neither a peak nor a shoulder corresponding to $S3$ could be resolved. In addition, no indication of the presence of $S3$ even above the Fermi level can be seen in any of the ARPES data recorded at room temperature (see Fig.\ \ref{fig:fits}-e 2f and, .\ \ref{fig:GamaMbandstrcuture} \cite{Epaps}).
On the other hand, comparing the ARPES results in which $S3$ has been detected, a certain inconsistency about $S3$ can be observed: A) $S3$ has different dispersions along the $\overline{\Gamma{\textit{M}}}$ direction in the literature \cite{hsieh_topological_2008,hsieh_observation_2009,nishide_direct_2010,nakamura_topological_2011,hirahara_topological_2010,guo_evolution_2011}. B) The reported energy separation between $S1$ and $S3$ at $k_x \approx 0.67$ $\text\AA^{-1}$ has different values ranging from $11$ to $45$\,meV \cite{hsieh_topological_2008,hsieh_observation_2009,nishide_direct_2010,nakamura_topological_2011,hirahara_topological_2010,guo_evolution_2011}. Hence, in contrast to $S1$ and $S2$, the detection of $S3$ seems to be not easily reproducible and to be more sporadic than systematic. However, the presence of $S3$ is not the ultimate proof of the non-trivial topology of the insulating BiSb alloy. \textit{Ab initio} and tight-binding calculations describe the topological phase with only $S1$ and $S2$ \cite{zhang_electronic_2009,teo_surface_2008}. Furthermore, the observed $S3$ in the experiment has not been related to the non-trivial topology but to surface imperfections \cite{zhang_electronic_2009,nakamura_topological_2011}. $S3$ can originate, for example, from locally different surface terminations \cite{zhu_topological_2014-1}. Thus, the uncontrolled damage to the surface caused by crystal cleaving can explain the sporadic character of $S3$. In this regard, the results presented here from \textit{in situ} grown films with comparatively reduced surface damage are closer to the realistic representation of the topological insulator $\text{Bi}_{1-x}\text{Sb}_{x}$.

With the absence of $S3$ in the non-trivial topological phase, $S1$ can not hybridize with $S3$, but merges into the conduction band near the $\overline{\textit{M}}$-point. The number of crossings with the Fermi level is still odd (five) \cite{zhang_electronic_2009,nakamura_topological_2011}. It ensues that the electron pocket at $\overline{\textit{M}}$ does not enclose the $\overline{\textit{M}}$-point in the Fermi surface \cite{zhang_electronic_2009,hsieh_topological_2008,hsieh_observation_2009}. In Fig.\ \ref{fig:GamaMbandstrcuture}, the measured Fermi surfaces of $\text{Bi}_{0.91}\text{Sb}_{0.09}$ and $\text{Bi}_{0.89}\text{Sb}_{0.11}$, which belong to the TI region are plotted. Closing electron-pockets contours just before the $\overline{\textit{M}}$-point can be discerned, especially for $x=0.09$. For $x=0.11$, the size of the electron pocket is smaller (Fig.\ \ref{fig:fits}-c) preventing us from resolving its contour outline. On the corresponding experimental surface band structures along $\overline{\Gamma{\textit{M}}}$ shown in the bottom panels, the intensity of $S1$ and $S2$ vanishes at $\overline{\textit{M}}$ (Fig.\ \ref{fig:GamaMbandstrcuture}). The red dashed lines are a guide to the eye. In addition, also the experimental band structure for $\text{Bi}_{0.7}\text{Sb}_{0.3}$ in Fig.\ \ref{fig:GamaMbandstrcuture} clearly shows a closed contour of the electron pocket and $S1$ merging into the conduction band before $\overline{\textit{M}}$. Consequently, the experimental surface band structure of the topologically non-trivial phase of $\text{Bi}_{1-x}\text{Sb}_{x}$(111) is in agreement with the theoretical modeling presented in Ref.\ \onlinecite{zhang_electronic_2009}.


We conclude that the defect-reduced $\text{Bi}_{1-x}\text{Sb}_{x}$(111) surface bears only two surface states $S1$ and $S2$ regardless of Sb concentration $x$. We schematically present the evolution of the surface states $S1$ and $S2$ as a function of Sb content $x$ in Fig.\ \ref{fig:Scenario}. For pure Bi(111) (Fig.\ \ref{fig:Scenario}-a), $S1$ has been considered to connect to the valence band at $\overline{\textit{M}}$ \cite{NB}. It switches connection from the valence band to the conduction band near $\overline{\textit{M}}$ at the topological transition ($x=0.04$) (Fig.\ \ref{fig:Scenario}-b). The surface state bands $S1$ and $S2$ further adapt to the energy shift of the bulk bands as Sb content increases. They smoothly evolve from a Bi(111)-like dispersion to the characteristic Sb(111)-like band structure (Fig.\ \ref{fig:Scenario}-c). This adaptation is most visible around $\overline{\Gamma}$ within $0.13 \lesssim x\lesssim 0.28$. However, it is not until $x\approx0.6$ that $S1$ and $S2$ become broad and faint near $\overline{\textit{M}}$ indicating  convergence to the Sb(111) band structure (Fig.\ \ref{fig:Scenario}-d) \cite{hsieh_observation_2009,hsieh_direct_2010}.

In summary, following an optimized method to grow high quality $\text{Bi}_{1-x}\text{Sb}_{x}$(111) films, we have investigated the evolution of the surface states of the system by a variation of $x$ from $0$ to $0.6$ using ARPES. Around $\overline{\Gamma}$ the ARPES data show a gradual evolution of the surface band structure from Bi(111) towards Sb(111). The previously reported third surface band near $\overline{\textit{M}}$ could not be detected in the topologically insulating phase here. We find our results of the experimental $\text{Bi}_{1-x}\text{Sb}_{x}$(111) surface state band structure to agree with available theoretical predictions, which identify the crystal as topologically non-trivial with two surface states.

We acknowledge stimulating discussions with A. Schnyder. H.\ M.\ B.\ acknowledges funding from the Deutsche Forschungsgemeinschaft (DFG). C.\ R.\ A.\ acknowledges funding from the Emmy-Noether-Program of the Deutsche Forschungsgemeinschaft (DFG).


\end{document}